\documentclass[superscriptaddress,prl,twocolumn,amsmath,amssymb,floatfix,showpacs]{revtex4}
\usepackage{graphicx}
\usepackage{bm}
\usepackage{amssymb}
\usepackage{color}
\usepackage{epsfig}
\usepackage{subfigure}
\usepackage{draftcopy}

\newcommand{\be}{\begin{equation}}
\newcommand{\ee}{\end{equation}}
\begin{document}

\title {Huge (but finite) time scales in slow relaxations: beyond simple aging}


\author{Ariel Amir}
\affiliation { Department of Condensed Matter Physics, Weizmann Institute of Science, Rehovot, 76100, Israel}
\affiliation { Department of Physics, Harvard University, Cambridge, MA 02138, USA}
\author{Stefano Borini}
\affiliation {Electromagnetic Division, INRIM, Strada delle Cacce 91, I-10135 Torino, Italy}
\author{Yuval Oreg}
\author{Yoseph Imry}
\affiliation { Department of Condensed Matter Physics, Weizmann Institute of Science, Rehovot, 76100, Israel}

\begin {abstract}
Experiments performed in the last years demonstrated slow relaxations and aging in the conductance of a large variety of materials. Here, we present experimental and theoretical results for conductance relaxation and aging for the case-study example of porous silicon. The relaxations are experimentally observed even at room temperature over timescales of hours, and when a strong electric field is applied for a time $t_w$, the ensuing relaxation depends on $t_w$. We derive a theoretical curve and show that all experimental data collapse onto it with a single timescale as a fitting parameter. This timescale is found to be of the order of thousands of seconds at room temperature. The generic theory suggested is not fine-tuned to porous silicon, and thus we believe the results should be universal, and the presented method should be applicable for many other systems manifesting memory and other glassy effects.

\end {abstract}



 \maketitle

For a large variety of systems in nature, the relaxation back to equilibrium depends strongly on the system's history. Examples of such behavior are abundant in the magnetic properties of spin-glasses \cite{vincent, spin_glasses, horacio}, the mechanical properties of polymers and biological material \cite{struik, plants, hairgel}, and in the behavior of colloids near the jamming transition \cite{aging_colloidal}. Understanding the aging properties is important for applied physics and engineering, as well as for theoretical physics, as this phenomenon is interrelated to the physics of glasses.

In the last few decades much progress has been made in understanding aging behavior in these broad range of physical systems, both theoretically and experimentally \cite{aging_models}. In one of the common aging protocols, the system is perturbed out of equilibrium for a time $t_w$, the waiting-time. A time $t$ after the perturbation is switched off, a physical observable in the system is measured. It was found that for various systems, the signal depends only on the ratio $t/t_w$, which is referred to as 'full' or 'simple' aging. Examples of full aging exist for a diversity of physical systems, such as  electron glasses \cite{zvi, Grenet, amir_review, popovic}, spin-glasses \cite{spinglass_kenning}, vortices in superconductors \cite{du} and granular systems \cite{granular_aging}. Nontheless, full aging is not generic. In some cases, a heuristic form of scaling is used $f(t/t_w^\mu)$ \cite{vincent2}, which was recently given theoretical support \cite{subaging}. Many systems, however, show more complex behavior, which is the case discussed here.

In this work, we study, experimentally and theoretically, the aging properties of a system of porous silicon, whose conductance exhibits aging behavior over timescales ranging from seconds to days.  It should be emphasized that the theoretical model does not contain ingredients which are specific to porous silicon, and should be universally applicable to a broad range of complex systems whose dynamics is governed by thermal activation or equivalent mechanisms. While some of the measurements appear to be close to full aging, clear and significant systematic deviations from it arise when the involved experimental timescales increase. This does not conform to the $f(t/t_w^\mu)$ scaling form mentioned above. The results suggest that the system must have unique, system-dependent finite timescale, characterizing the breakdown of the full aging regime. Here, we consider a model that predicts a particular form of aging, and demonstrate how it fully accounts for the complex form of aging in porous silicon. We succeed in accurately deducing a timescale characterizing the system, and find that, experimentally, it is sensitive to the temperature as well as to the time since the sample has been manufactured,\emph{ i.e.}, its 'age' in the usual sense.

The structure of the manuscript is as follows. We first present the porous silicon system and the experimental protocol. We proceed by outlining the physical model, which was used by some of us earlier in the context of electron glasses, and put forth the prediction for the aging behavior, including the explicit dependence on the system dependent timescale. This timescale has the physical significance of the reciprocal of the lower cutoff of the underlying relaxation distribution.  Finally, we show the excellent agreement of experimental data and theory, both for the 'excitation' and 'relaxation' parts of the experimental protocol, which will be presently defined.

\emph{Experimental System.-} There has been much interest in the physics of porous silicon, due to its potential in applications such as biosensors \cite{silicon_biosensors} and its remarkable optical properties \cite{silicon_photolum}, and due to the high controllability of its fabrication.
Mesoporous silicon (MesoPS) consist of a disordered network of interconnected silicon nanocrystals. For the samples studied here the size of the nanocrystals is of the order of tens of nanometers \cite{silicon_photolum}.  The samples are fabricated by electrochemical etching of heavily doped p-type silicon in hydrofluoric acid solution. The linear dimensions of the samples are of the order of tens of microns. Since the obtained mesoPS layers are not thin, the role of contact resistance and of the silicon substrate in the electrical characterization is negligible. The electronic conduction mechanism in this material is not fully understood. Previous studies have suggested that the conduction is limited by various narrow channels of porous silicon, where due to the presence of nearby trapped charges the electronic passage is severely suppressed \cite{Lehmann1995_}. However, the conduction dependence on temperature in some cases is reminiscent of variable-range-hopping \cite{Zimin}, which may point out to existence of other possible conduction mechanisms at lower temperatures. Full details of the experimental system are presented in \cite{borini_prb2007}.

The I-V characteristics of the porous silicon samples are empirically found to be superohmic. 
This implies that when pushed out-of-equilibrium by an electric field, the time dependent conductance (\emph{i.e.}, the current to voltage ratio) must increase to its higher out-of-equilibrium value. Remarkably, it does so very slowly, over the course of hours.

%

\begin{figure}[b!]
\includegraphics[width=0.4\textwidth]{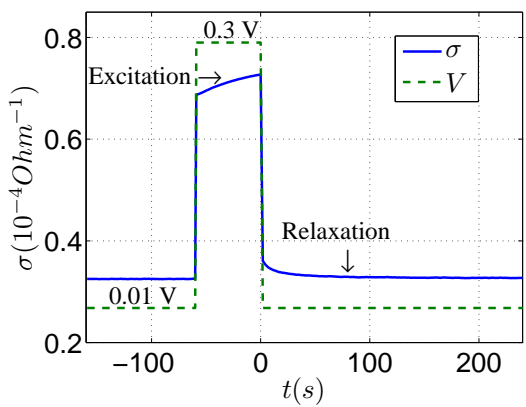}
\caption{ Measurement of time-dependent conductance. The electric field exerted on the system, $F$, is changed from an initial value to a larger value for a time $t_w$, as shown in the figure. Upon the increase of field, the conductance increases monotonically, attempting to reach a new, higher steady-state value, but does not saturate on the experimental timescales. Upon the restoration of the electric field to its initial value, the conductance relaxes in a form different from exponential.\label{protocol} }
\end{figure}

The protocol used in the experiment is as follows. First, the sample is let to relax for a long time, of the order of several hours, in a constant electric field $F$, presumably reaching very close to its steady-state. The value of the field is then raised, for a time $t_w$, which can be anything from seconds to a few hours. The initial value of the electric field is now restored, and the system relaxes to its initial steady-state. The current is measured as a function of time, through the experiment. Fig. \ref{protocol} shows the results for a particular value of $t_w$.

\emph{Model.-} In \cite{amir_review}, it was argued that various, generic physical mechanisms may lead to a broad distribution of relaxation rates which is approximately: $P(\lambda) \sim 1/\lambda$. One of these, which we believe to be the relevant one here, relies on thermally activated processes. Indeed, if we assume such processes, $\lambda \sim e^{-U/T}$, and a distribution of barriers which is approximately uniform between a lower and upper cutoff, $P(U) \approx C$, we obtain the above distribution of relaxation rates (also between a lower and upper cutoff, $\lambda_{\rm{min}}$ and $\lambda_{\rm{max}}$). This simple physical reasoning is at the heart of seminal works on 1/\emph{f} noise \cite{VanDerZiel}.

In the following we will adopt the aging picture described in Ref. [\onlinecite{amir_aging}], with an important difference: here, the cutoffs of the distribution \emph{will} play a major role, and we will \emph{not} assume that we are in the asymptotic regime where they do not govern the physics, namely $1/\lambda_{max} \ll t \ll 1/\lambda_{min}$ will not be valid here. Let us review the basic mechanism: when the external perturbation (voltage in our case) is applied, the slow modes are partially excited, and the conductance of the system increases to a larger out-of-equilibrium saturation value. In complete analogy to charging a capacitor, the excitation of a single mode with characteristic frequency $\lambda$ is $[1-e^{-\lambda t}]$. Using the above distribution for the slow modes, we thus obtain:

\be \delta \sigma^{ex} (\hat{t}) \propto \int_{\lambda_{\rm{min}}}^{\lambda_{\rm{max}}} \frac{1}{\lambda}[1-e^{-\lambda \hat{t}}] d\lambda, \ee with $\hat{t}$ the time elapsed from the application of the perturbation.

Thus in the excitation regime:
\be \delta \sigma^{ex} (\hat{t}) \propto A+E_1[\lambda_{\rm{max}}\hat{t}]-E_1[\lambda_{\rm{min}}\hat{t}] \label{excitation_theory},\ee

with $A=\log(\lambda_{\rm{max}}/\lambda_{\rm{min}})$, and $E_1$ the exponential integral function.
As long as we do not consider too short times, \emph{i.e.}, as long as $\hat{t} \gg 1/\lambda_{max}$, we can neglect the term $E_1[\lambda_{\rm{max}}\hat{t}]$.

Similarly, in the aging regime (see Fig. \ref{protocol}), each of the modes relaxes exponentially with rate $\lambda$. Therefore the excess conductance is:

\be \delta \sigma^{ag} (t,t_w) \propto \int_{\lambda_{\rm{min}}}^{\lambda_{\rm{max}}} \frac{1}{\lambda}[1-e^{-\lambda t_w}]e^{-\lambda t} d\lambda. \ee

Which can be approximated for times which are longer than $1/\lambda_{\rm{max}}$ as:

\be \delta \sigma^{ag} (t,t_w) \propto E_1[\lambda_{\rm{min}}t]-E_1[\lambda_{\rm{min}}(t+t_w)]. \label{aging_theory} \ee

Were we to assume that the experimental timescales are much shorter than the reciprocal cutoff $\lambda_{\rm{min}}$, we could approximate the exponential integral function by a logarithm. Under that approximation, the cutoff drops out of the final expression for the excess conductance, yielding the form $\rm{log}(1+t_w/t)$. This is the case for electron glasses, which has been experimentally observed for a variety of systems \cite{amir_aging}. The remarkable thing about the porous silicon system, is that the timescale associated with the cutoff $\lambda_{\rm{min}}$ is long enough such that one can measure slow relaxations over the course of hours, but, it is still experimentally measurable, and its finite value \emph{must} be considered in order to understand the \emph{a-priori} complex aging behavior. In fact, the deviations from full aging in this case arise precisely from the finite value of $\lambda_{\rm{min}}$ (and will disappear as the experimental timescales will become shorter and shorter).

\begin{figure}[b!]
\includegraphics[width=0.45 \textwidth]{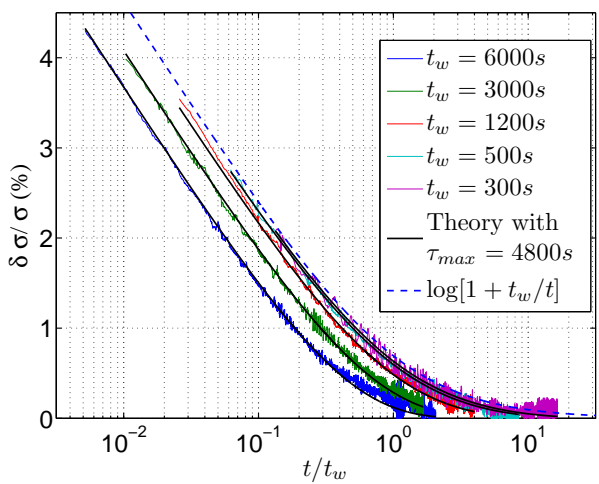}
\caption{Experimental results and theory compared, for a specific porous silicon sample, measured at $T=299K$. The x axis shows the time divided by the waiting-time $t_w$, which is different for each of the curves. The y axis shows the excess conductance from its equilibrium value. The experiment was performed for waiting-time values of $t_w=300s,500s,1200s, 3000s, 6000s$. Full aging corresponds to the case were there would be complete data collapse of all curves. Here, while the curves for the smaller values of $t_w$ are close to complete data collapse, the curves corresponding to the larger waiting-times are significantly off from it. The solid lines correspond to the theoretical form of Eq. (\ref{aging_theory}), with a single fitting parameter $\lambda_{\rm{min}}$, \emph{\textbf{for all of the five curves.}} \label{aging} }
\end{figure}

\emph{Results.-} Fig. \ref{aging} compares Eq. (\ref{aging_theory}) and the experimental results. $\lambda_{\rm{min}}$ is taken as a fitting parameter. The excellent agreement using a single fitting parameter to data collapse five curves, shows that the theoretical framework suggested here is adequate. It thus allows us to associate a \emph{single} timescale $\tau_{max} \equiv 1/\lambda_{\rm{min}}$ with each porous silicon sample (which is typically of the order of thousands of seconds), which has the physical meaning of being the longest relaxation time existing in the system.

\begin{figure}[b!]
\includegraphics[width=0.4\textwidth]{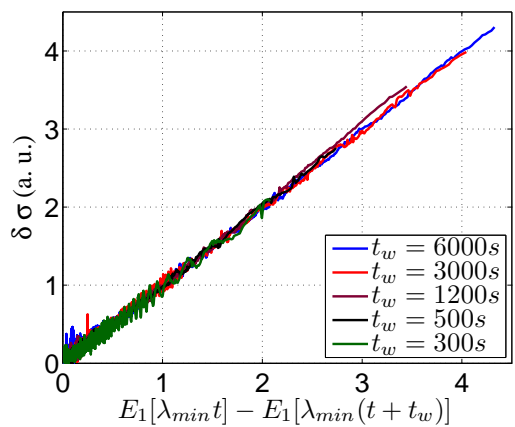}
\caption{The same data of Fig. \ref{aging} is analyzed in a different way: the time (x axis) is rescaled according to Eq. (\ref{aging_theory}), leading to a linear dependence of the excess conductance. For electron glasses \cite{amir_aging}, rescaling according to $\log(1+t_w/t)$ would be adequate, and one would not be able to use this procedure to deduce a timescale. Here, the exponential integral function is necessary, and not only its asymptotic form. \label{linearized} }
\end{figure}

Fig. \ref{linearized} shows the analysis of the same data set in another way, by rescaling time according to Eq. (\ref{aging_theory}), leading to a linear dependence of the excess conductance on the rescaled variable. i.e., we present a parametric plot where the $y$-axis is the measured conductance and the $x$-axis is given by Eq. (\ref{aging_theory}).

\begin{figure}[b!]
\includegraphics[width=0.4 \textwidth]{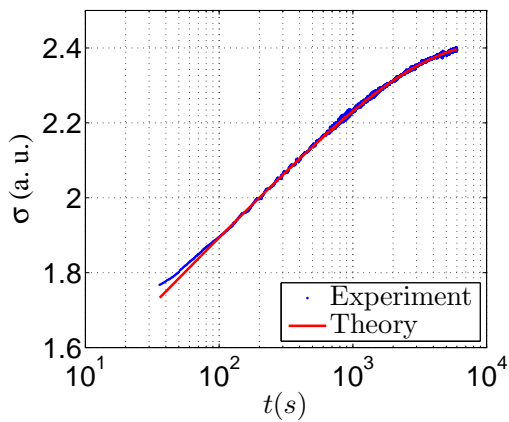}
\caption{ The change in conductance during the `excitation' phase, when the system is driven out-of-equilibrium, for the same sample and conditions of Figs. \ref{aging} and \ref{linearized}. In accordance to Eq. (\ref{excitation_theory}), the rise in conductance is described by an exponential integral function, with a single fitting parameter $\lambda_{\rm{min}}$. Its value was determined by the aging characteristics (see Fig. \ref{aging}), and therefore in the above figure $\lambda_{\rm{min}}$ is \emph{not} a fitting parameter. The good agreement between the theory and experiment prove that the excitation of the conductance results from the same physical origin as the relaxation. \label{excitation_fig} }
\end{figure}

Having determined the timescale $\tau_{max}$ from the aging experiments, we test its applicability also during the excitation step, where the conductance is expected to increase as an exponential integral function, as described by Eq. (\ref{excitation_theory}). Fig. (\ref{excitation_fig}) shows the good agreement between experiments and theory, where now $\lambda_{\rm{min}}$ \textbf{is no longer a fitting parameter}. This is consistent with the suggested framework, in which the modes which are excited during the application of the electric field are the same ones which relax during the aging phase.

So far we have established the physical significance of a timescale characterizing the system. Having done that, one can now investigate the dependence of $\tau_{max}$ on the system's parameters (\emph{e.g}: temperature), and learn about the underlying physical mechanisms.
For the case of porous silicon, we have found that $\tau_{max}$ is sensitive to the temperature, which is consistent with the thermal activation mechanism. For the sample measured, upon changing the temperature from 300K to 280K the timescale $\tau_{max}$ changed from about 800 seconds to more than 5000 seconds (it should be noted that the measurements of Figs. \ref{aging}-\ref{excitation_fig} were done on a different sample, giving different timescales). Using measurements at several different temperatures would allow one to find the cutoff of the barrier distribution, \emph{i.e.}, the deepest traps available. Moreover, we have found that leaving the samples exposed to air for a long period (of the order of several weeks) substantially influences this timescale: the value of $\tau_{max}$ was 160 seconds for a freshly made sample, but grew to 4800s for a two months old sample. This is another form of `aging', which is maybe closer to the everyday use of this terminology: as the sample grows older, its typical timescale grows. Here, it means, remarkably, that the effective traps in the porous silicon are not static in time, but are `deepened' as time progresses. This is a further process, occurring on much longer timescales than those involved in the experimental protocol, which justifies the analysis with a constant in time $\tau_{max}$. We believe our analysis demonstrates the power in the method, which allows one to gain insights into the microscopics of porous silicon, through the macroscopic measurements of the conductance's slow relaxations. For example, the determination of the system's characteristic timescale was recently discussed in the context of electron glasses \cite{Grenet_timescale}, and the method presented here might prove useful also in that case.

\emph{Summary.-} We studied slow relaxations and aging in a system of porous silicon, which is well suited for this purpose since it supports relaxation modes which are extremely long, yet it has a finite, measurable timescale $\tau_{max}$. We have shown that one can understand quantitatively the aging behavior of the conductance in the system, and data collapse the measurements from five different experiments using a single (sample dependent) fitting parameter, the timescale $\tau_{max}$. The underlying theory used was not particular to porous silicon, and certain limits of it, where the aging is independent of the timescale $\tau_{max}$, have been shown to describe other systems \cite{amir_aging}, suggesting that the results are robust. We have shown that $\tau_{max}$ is temperature sensitive, consistent with the picture of thermal activation over large energetic barriers in the systems. Since the theoretical model does not contain any ingredients particular to porous silicon, we believe these intriguing phenomena should be applicable to various other systems where deviations from full aging are observed, in material science and in other glassy systems.

We acknowledge the hospitality of KITP at Santa Barbara where part of this research was performed, and the participants of the Electron Glass Workshop for stimulating discussions. This work was also supported by a BMBF DIP grant as well as by ISF and BSF grants and the Center of Excellence Program. YI was also supported by a continuing Humboldt award.


\begin{thebibliography}{10}

\bibitem{vincent}
V. Dupuis {\it et~al.}, Pramana J. of Phys. {\bf 64},  1109  (2005).

\bibitem{spin_glasses}
J.~P. Bouchaud, L. Cugliandolo, J. Kurchan, and M. Mezard,  in {\em Spin
  Glasses and Random Fields}, edited by A.~P. Young (World Scientific,
  Singapore, 1998).

\bibitem{horacio}
H.~E. Castillo, C. Chamon, L.~F. Cugliandolo, and M.~P. Kennett, Phys. Rev.
  Lett. {\bf 88},  237201  (2002).

\bibitem{struik}
L.~C.~E. Struik, {\em Physical Aging in Amorphous Polymers and Other Materials}
  (Elsevier, Oxford, 1978).

\bibitem{plants}
D.~S. Thompson, J. Exp. Bot. {\bf 52},  1291  (2001).

\bibitem{hairgel}
A. Shahin and Y.~M. Joshi, Phys. Rev. Lett. {\bf 106},  038302  (2011).

\bibitem{aging_colloidal}
L. Cipelletti and E. R. Weeks, Glassy dynamics and dynamical heterogeneity in
  colloids, arXiv:1009.6089 (2010).

\bibitem{aging_models}
P. Sibani and K.H. Hoffmann, Phys. Rev. Lett. {\bf 63}, 2853 (1989); J. P.
  Bouchaud, J. Phys. I (France) {\bf 2}, 1705 (1992); L.F. Cugliandolo and J.
  Kurchan, Phys. Rev. Lett. {\bf 71}, 173 (1993); L.F. Cugliandolo, J. Kurchan,
  and F. Ritort, Phys. Rev. B {\bf 49}, 6331 1994); S. Franz and J. Hertz,
  Phys. Rev. Lett. {\bf 74}, 2114 (1995); C. Monthus and J. P. Bouchaud, J.
  Phys. A: Math. Gen. {\bf 29}, 3847 (1996); E. Marinari, G. Parisi, and D.
  Rossetti, Eur. Phys. J. B {\bf 2}, 495 (1998).

\bibitem{zvi}
A. Vaknin, Z. Ovadyahu, and M. Pollak, Phys. Rev. Lett. {\bf 84}, 3402 (2000);
  V. Orlyanchik and Z. Ovadyahu, Phys. Rev. Lett. {\bf 92}, 066801 (2004); Z.
  Ovadyahu and M. Pollak, Phys. Rev. B {\bf 68}, 184204 (2003); Z. Ovadyahu,
  Phys. Rev. B {\bf 73}, 214204 (2006).

\bibitem{Grenet}
T. Grenet, Eur. Phys. J. B {\bf 32}, 275 (2003); T. Grenet, Phys. Stat. Sol.
  (c) {\bf 1}, 9 (2004); T. Grenet, J. Delahaye, M. Sabra, and F. Gay, Eur.
  Phys. J B 56, 183 (2007).

\bibitem{amir_review}
A. Amir, Y. Oreg, and Y. Imry, Annu. Rev. Condens. Matter Phys. {\bf 2},  235
  (2011).

\bibitem{popovic}
J. Jaroszynski and D. Popovic,
Phys. Rev. Lett. {\bf 99}, 216401 (2007).

\bibitem{spinglass_kenning}
G.~F. Rodriguez, G.~G. Kenning, and R. Orbach, Phys. Rev. Lett. {\bf 91},
  037203  (2003).

\bibitem{du}
X. Du {\it et~al.}, Nat. Phys. {\bf 3},  111  (2007).

\bibitem{granular_aging}
N. Kurzweil and A. Frydman, Phys. Rev. B {\bf 75},  020202  (2007).

\bibitem{vincent2}
E. Vincent, Lecture notes in physics {\bf 716},  7  (2007).

\bibitem{subaging}
P. Sibani and G.~G. Kenning, Phys. Rev. E {\bf 81},  011108  (2010).

\bibitem{silicon_biosensors}
A. Jane, R. Dronov, A. Hodges, and N. Voelcker, Trends in Biotechnology {\bf
  27},  230  (2009).

\bibitem{silicon_photolum}
A.~G. Cullis, L.~T. Canham, and P.~D.~J. Calcott, J. Appl. Phys. {\bf 82},  909
   (1997).

\bibitem{Lehmann1995_}
V. Lehmann, F. Hofmann, F. M{\"{u}}ller, and U. Gr{\"{u}}ning, Thin Solid Films
  {\bf 255},  20   (1995), european Materials Research Society 1994 Spring
  Conference, Symposium F: Porous Silicon and Related Materials.

\bibitem{Zimin}
S.~P. Zimin, Semiconductors {\bf 40},  1350  (2006).

\bibitem{borini_prb2007}
S. Borini, L. Boarino, and G. Amato, Phys. Rev. B {\bf 75},  165205  (2007).

\bibitem{VanDerZiel}
A.~V.~D. Ziel, Physica {\bf 16},  359   (1950).

\bibitem{amir_aging}
A. Amir, Y. Oreg, and Y. Imry, Phys. Rev. Lett. {\bf 103},  126403  (2009).

\bibitem{Grenet_timescale}
T. Grenet and J. Delahaye, arXiv:1105.0984 (2011).

\end{thebibliography}
\end{document}